\documentclass[twocolumn,showpacs,aps,prd]{revtex4-1}
\usepackage{graphicx}
\usepackage{float}
\usepackage{epsfig,graphics,subfigure,psfrag,amsmath,amssymb}
\usepackage{lineno}
\usepackage{dcolumn}
\usepackage{bm}
\usepackage{overpic}
\usepackage{xspace}
\usepackage{rotating}
\usepackage{epstopdf}
\usepackage{makecell}
\usepackage{multirow}
\usepackage[bookmarksnumbered, pdfstartview=FitH,colorlinks,urlcolor=blue, citecolor=blue,linkcolor=blue] {hyperref}

\begin{document}
\title{\bf \boldmath Observation of the decay $D^+_s\to \omega\pi^+\eta$}

\author{
\begin{small}
\begin{center}
M.~Ablikim$^{1}$, M.~N.~Achasov$^{12,b}$, P.~Adlarson$^{72}$, M.~Albrecht$^{4}$, R.~Aliberti$^{33}$, A.~Amoroso$^{71A,71C}$, M.~R.~An$^{37}$, Q.~An$^{68,55}$, Y.~Bai$^{54}$, O.~Bakina$^{34}$
% ...
	\\
\vspace{0.2cm}
(BESIII Collaboration)\\
	\vspace{0.2cm}{\it
$^{1}$ Institute of High Energy Physics, Beijing 100049, People's Republic of China\\
% ...
$^{4}$ Bochum  Ruhr-University, D-44780 Bochum, Germany\\
% ...
$^{12}$ G.I. Budker Institute of Nuclear Physics SB RAS (BINP), Novosibirsk 630090, Russia\\
% ...
$^{33}$ Johannes Gutenberg University of Mainz, Johann-Joachim-Becher-Weg 45, D-55099 Mainz, Germany\\
$^{34}$ Joint Institute for Nuclear Research, 141980 Dubna, Moscow region, Russia\\
% ...
$^{37}$ Liaoning Normal University, Dalian 116029, People's Republic of China\\
% ...
$^{54}$ Southeast University, Nanjing 211100, People's Republic of China\\
$^{55}$ State Key Laboratory of Particle Detection and Electronics, Beijing 100049, Hefei 230026, People's Republic of China\\
% ...
$^{68}$ University of Science and Technology of China, Hefei 230026, People's Republic of China\\
% ...
$^{71}$ University of Turin and INFN, (A)University of Turin, I-10125, Turin, Italy; (B)University of Eastern Piedmont, I-15121, Alessandria, Italy; (C)INFN, I-10125, Turin, Italy\\
$^{72}$ Uppsala University, Box 516, SE-75120 Uppsala, Sweden\\
% ...
$^{b}$ Also at the Novosibirsk State University, Novosibirsk, 630090, Russia\\
% ...
	}
\end{center}
\vspace{0.4cm}
\end{small}
}

\begin{abstract}
  Using 7.33 fb$^{-1}$ of $e^+e^-$ collision data collected by the BESIII detector at center-of-mass energies between 4.128 and 4.226~GeV, we observe for the first time the decay $D^{\pm}_s\to \omega\pi^{\pm}\eta$ with a statistical significance of 7.6$\sigma$.  The measured branching fraction of this decay is $(0.54\pm0.12\pm0.04)\%$, where the first uncertainty is statistical and the second is systematic.
\end{abstract}

\maketitle
\section{INTRODUCTION}
Hints of deviations from the Standard Model have been seen in the measurements of the decays of bottom and charm hadrons.
In particular, the ratio $R(D^{*}) \equiv \mathcal{B}(B\to D^{*}\tau^+\nu_{\tau})/\mathcal{B}(B\to D^{*}\mu^+\nu_{\mu}) = 0.295\pm0.010\pm0.010$ provided by the Heavy Flavor Averaging Group~\cite{HFLAV:2019otj} differs from the Standard Model prediction $0.258\pm0.005$ by 2.5 standard deviations, 
showing a possible violation of the lepton flavor universality~(LFU). 
The $\mathcal{B}$ is defined as the branching fraction~(BF).
The current data taking at $B$-factories and LHCb will increase the data sample, hence reducing the statistical uncertainty, while an improvement of the systematic uncertainty is also necessary.
The $R(D^*)$ measurement in the LHCb experiment~\cite{LHCb:2017smo, LHCb:2017rln} suffers from 
limited knowledge of the $D^+_s\to \pi^+\pi^+\pi^- X$ decay, where $X$ stands for all possible particle combinations, 
since the decay of $B^0\to D^{*-}D^+_s, D^+_s\to \pi^+\pi^+\pi^- X$ is the main background for the analysis of $B^0\to D^{*-}\tau^+\nu_{\tau}, \tau^+\to \pi^+\pi^+\pi^- X$.
Measurements of the BFs of hadronic $D^+_s$ decays including at least three charged pions in the final states offer key inputs to improve the precision of $R(D^*)$.  
This work focuses on the previously unobserved mode $D^+_s\to \omega \pi^+\eta$, where we require the decay $\omega\to\pi^+\pi^-\pi^0$ .  

In addition, hadronic $D_s^{\pm}$ decays probe the interplay of short-distance weak decay matrix elements and long-distance QCD effects in charm meson decays. 
But information is still limited for the $D_s^+$ case, where a large part of the hadronic BF is still unmeasured~\cite{Li:2021iwf}. 
The Cabibbo-favored hadronic $D^+_s$ decays mediated via a $c\to s W^+, W^+\to u\bar{d}$ transition and producing states with hidden strangeness have relatively large BFs.
According to Particle Data Group (PDG)~\cite{PDG}, the missing hadronic decays of $D_s^+$ with $\omega$ ($\eta$) in the final state contribute a fraction of $(1.5\pm1.6)\%~((3.0\pm3.1)\%)$. 
Among them, the $D_s^+\to \omega\pi^+\eta$ decay is predicted to have a relatively large decay rate~\cite{Gronau:2009vt}.
The CLEO collaboration~\cite{CLEO:2009nsf} claimed evidence for this mode with $\mathcal{B}(D^+_s\to \omega\pi^+\eta) = (0.85\pm0.54_{\rm stat.}\pm0.06_{\rm syst.})\%$, based on $4.9\pm2.9$ signal events using 0.586 fb$^{-1}$ of $e^+e^-$ collision data taken at the center-of-mass energy $\sqrt{s}$ = 4.17 GeV. 
In this paper, we present the first observation and the BF measurement of $D^+_s\to \omega\pi^+\eta$ decay using 7.33~fb$^{-1}$ of $e^+e^-$ collision data collected with the BESIII detector between $\sqrt{s} = 4.128$ and $4.226$~GeV. Charge conjugate states are implied throughout this paper.

\section{DETECTOR AND DATA SETS}
\label{sec:detector_dataset}
The BESIII detector~\cite{BESIII:2009fln} records symmetric $e^+e^-$ collisions provided by the BEPCII storage ring~\cite{Yu:2016cof}, in center-of-mass energies range from 2.0 to 4.95 GeV, with a peak luminosity of $1\times 10^{33}$ cm$^{-2}$s$^{-1}$ achieved at center-of-mass energy of 3.77 GeV.
The cylindrical core of the BESIII detector covers 93\% of the full solid angle and consists of a helium-based multilayer drift chamber~(MDC), 
a plastic scintillator time-of-flight system~(TOF), and a CsI(Tl) electromagnetic calorimeter~(EMC), 
which are all enclosed in a superconducting solenoidal magnet providing a 1.0~T magnetic field~\cite{Huang:2022wuo}. 
The solenoid is supported by an octagonal flux-return yoke with resistive plate counter muon identification modules interleaved with steel.
The charged-particle momentum resolution at 1~GeV/$c$ is $0.5\%$, and the ${\rm d}E/{\rm d}x$ resolution is $6\%$ for electrons from Bhabha scattering. 
The EMC measures photon energies with a resolution of $2.5\%$ ($5\%$) at 1~GeV in the barrel (end cap) region.
The time resolution in the TOF barrel region is 68~ps, while that in the end cap region is 110~ps. 
The end cap TOF system was upgraded in 2015 using multi-gap resistive plate chamber technology, providing a time resolution of 60~ps~\cite{etof1}. 
About $84\%$ of the data in this analysis benefits from the upgrade.

The data samples are organized into four groups, $\sqrt{s} =$ 4.128 and 4.157 GeV, 4.178 GeV, four energies from 4.189-4.219 GeV, and 4.226 GeV, acquired during the same year under consistent running conditions. 
The integrated luminosities at each energy are given in Table~\ref{energe}.
Since the cross section of $e^+e^-\to D^{*\pm}_sD_s^{\mp}$ production in $e^+e^-$ annihilation is about twenty times larger than the $e^+e^-\to D^+_sD^-_s$ one~\cite{CLEO:2008ojp}, 
the signal events discussed in this paper are selected from the process $e^+e^-\to D^{*\pm}_sD_s^{\mp}$.

\begin{table}[htbp]
 \centering
 \caption{The integrated luminosities ($\mathcal{L}_{\rm int}$) and the requirements on $M_{\rm rec}$ for various center-of-mass energies.
	 The definition of $M_{\rm rec}$ is given in Eq.~(\ref{eq:mrec}).
   The first and second uncertainties are statistical and systematic, respectively.}
 \begin{tabular}{c c c}
 \hline
	 $\sqrt{s}$ (GeV)~\cite{BESIII:2020eyu} &$\mathcal{L}_{\rm int}$ (pb$^{-1}$)~\cite{luminosities} & $M_{\rm rec}$ (GeV/$c^2$)\\
 \hline
	 4.128 &401.5   &[2.060, 2.150] \\
	 4.157 &408.7   &[2.054, 2.170] \\
	 4.178 &$3189.0\pm0.2\pm31.9$  &[2.050, 2.180] \\
	 4.189 &$570.0\pm0.1\pm2.2$ &[2.048, 2.190] \\
	 4.199 &$526.0\pm0.1\pm2.1$ &[2.046, 2.200] \\
	 4.209 &$572.1\pm0.1\pm1.8$ &[2.044, 2.210] \\
	 4.219 &$569.2\pm0.1\pm1.8$ &[2.042, 2.220] \\
	 4.226 &$1100.9\pm0.1\pm7.0$ &[2.040, 2.220] \\
  \hline
 \end{tabular}
 \label{energe}
\end{table}

Inclusive Monte Carlo~(MC) samples, 40 times larger than the data sets, are produced between $\sqrt{s} = 4.128$ and $4.226$ GeV with a {\sc geant4}-based~\cite{GEANT4:2002zbu} toolkit, which includes the geometric description of the BESIII detector and the detector response. 
These samples are used to determine the detection efficiencies and to estimate backgrounds.
The samples include the production of open charm processes, the initial-state radiation production of vector charmonium(-like) states and the continuum processes incorporated in {\sc kkmc}~\cite{Jadach:2000ir, Jadach:1999vf}. 
All particle decays are modelled with {\sc evtgen}~\cite{Lange:2001uf, EVTGEN2} using BFs either reported by the PDG~\cite{PDG}, 
when available, or otherwise estimated with {\sc lundcharm}~\cite{Chen:2000tv, LUNDCHARM2}. 
Final state radiation from charged final state particles is incorporated using {\sc photos}~\cite{PHOTOS}. 
The signal detection efficiencies and signal shapes are obtained with the signal MC samples. 
The signal decay $D^+_s\to\omega\pi^+\eta$ is generated with a phase-space distribution.
The MC samples of the decays $D^+_s\to\omega a_0(980)^+$ and $D^+_s\to\eta b_1(1235)^+$, used to estimate the systematic uncertainty of MC models, are generated with the SVS model, which describes the decay of a scalar meson to vector plus scalar mesons~\cite{Lange:2001uf, EVTGEN2}.

\section{METHODOLOGY}
The data samples were collected just above the $D_s^{*\pm}D_s^{\mp}$ 
threshold, such that the $D_s^{*\pm}D_s^{\mp}$ system is produced exclusively, 
without any additional hadrons.  
The tag method~\cite{MARK-III:1985hbd} was used, which allows to select clean signal samples, providing the opportunity to measure the absolute BFs of the hadronic $D^+_s$ meson decays. 
In the tag method, a single-tag (ST) candidate requires only one of the $D_{s}^{\pm}$ mesons to be reconstructed via a hadronic decay. 
In a double-tag (DT) candidate, both the $D_s^+$ and $D_s^-$ mesons are reconstructed, with the $D_{s}^{+}$ meson decaying to the signal mode $D_{s}^{+} \to \omega\pi^+\eta$ and the $D_{s}^{-}$ meson decaying to one of the tag modes. 
The decays $D_{s}^{-}\to K_{S}^{0}K^{-}$, $D_{s}^{-}\to K^{+}K^{-}\pi^{-}$ and $D^{-}_{s}\to K^{+}K^{-}\pi^{-}\pi^{0}$ are chosen as tag modes, based on the optimization of the figure of merit for DT yields, defined as $S/\sqrt{S+B}$. 
Here, $S$ and $B$ denote the signal and background yields estimated by the inclusive MC samples, respectively.

\begin{figure*}[!]
\begin{center}
    \includegraphics[width=5.9cm]{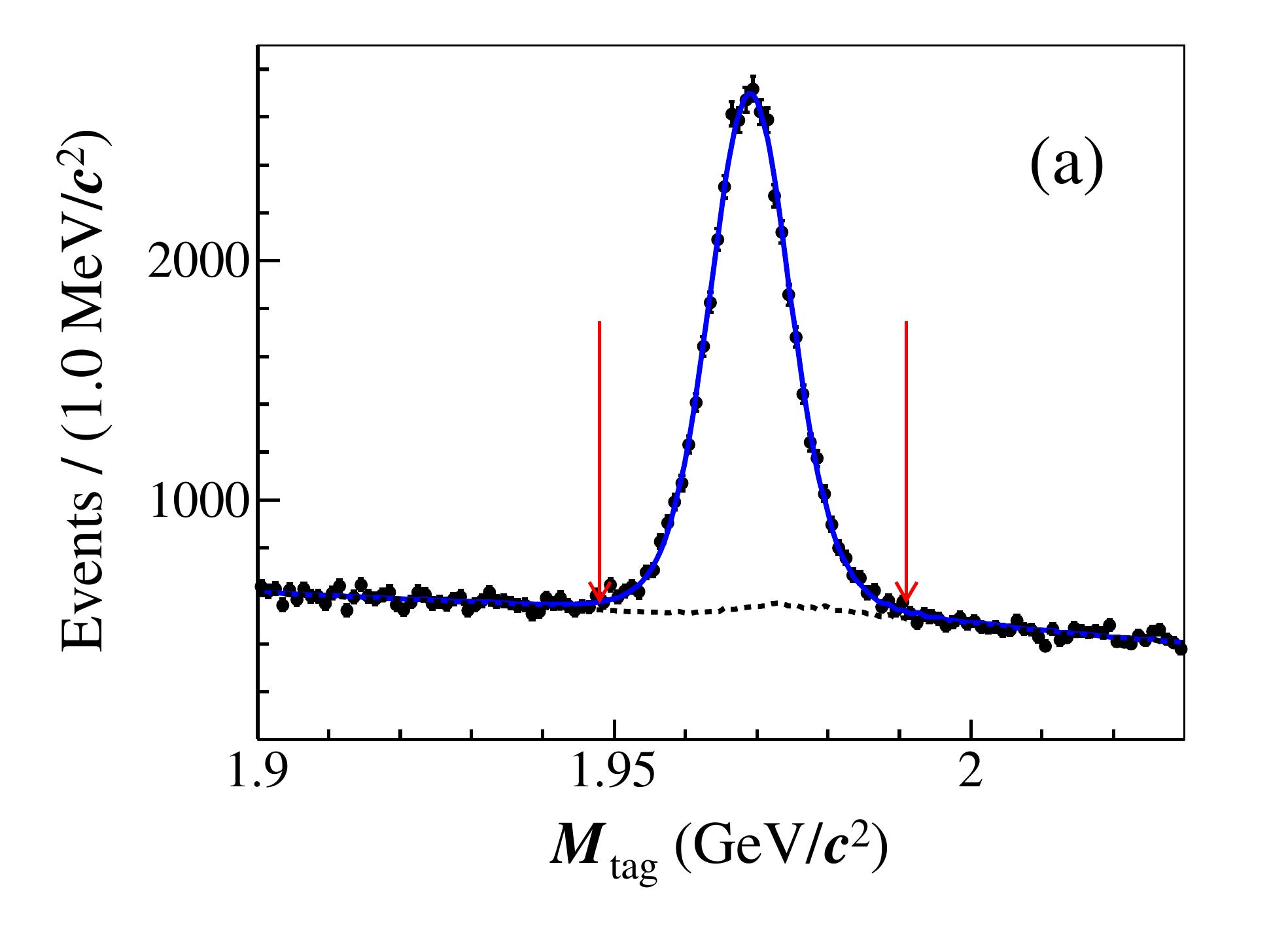}
    \includegraphics[width=5.9cm]{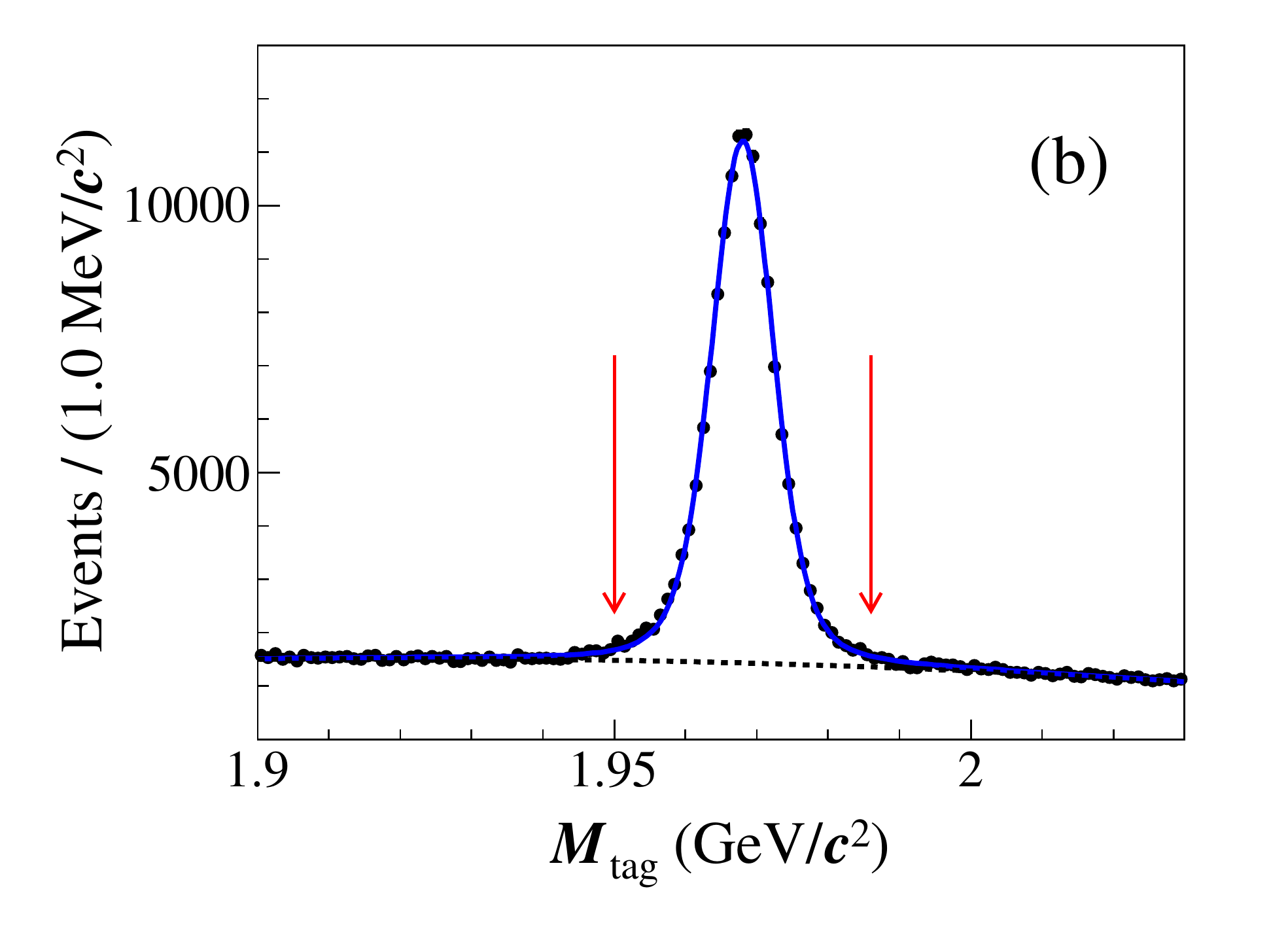}
    \includegraphics[width=5.9cm]{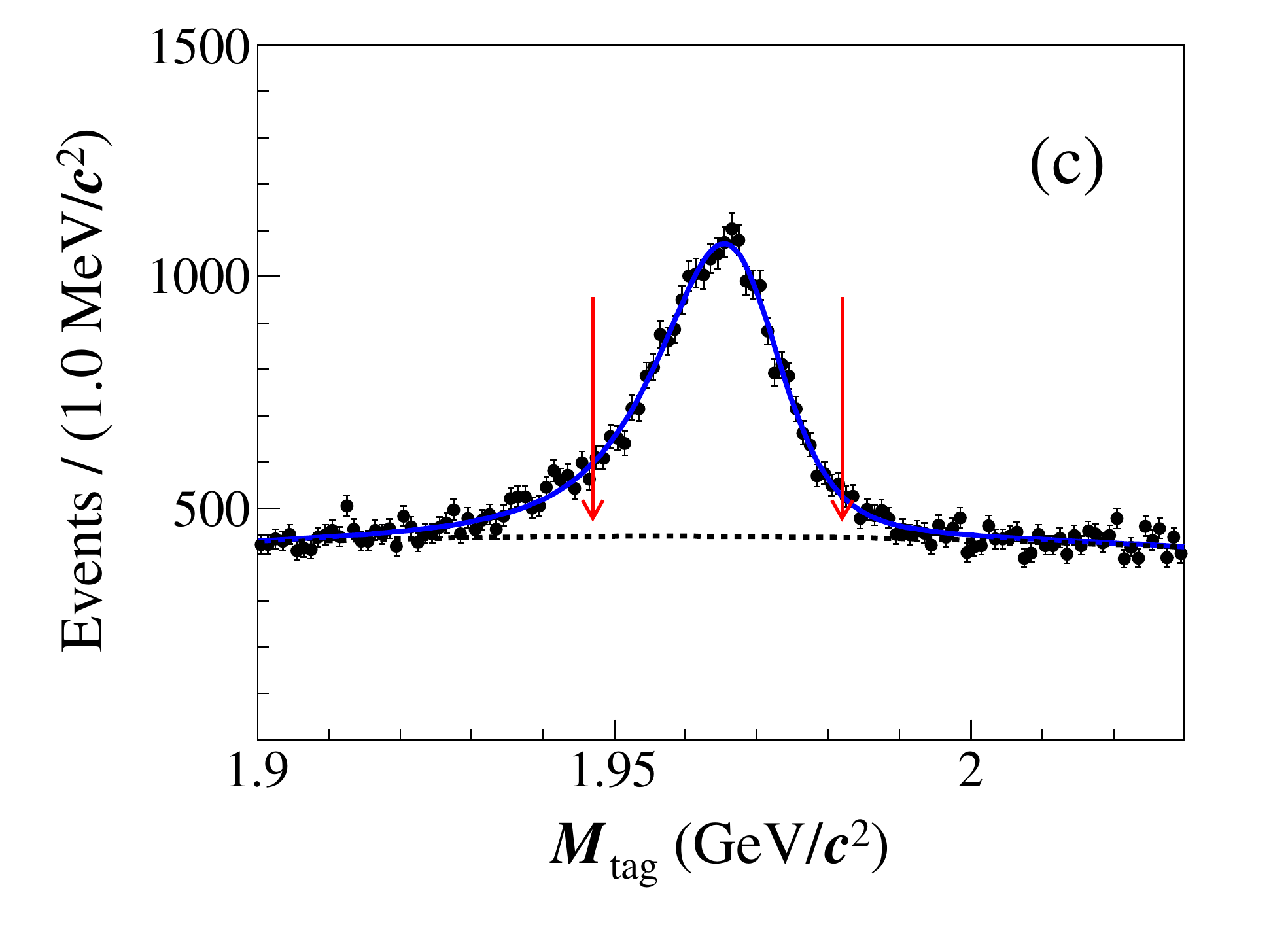}
	\caption{
	Fits to the $M_{\rm tag}$ distributions of the ST candidates for (a)~$D_s^-\to K_S^0K^-$, (b)~$D_s^-\to K^+K^-\pi^-$ and (c)~$D_s^-\to K^+K^-\pi^-\pi^0$ from the data sample at $\sqrt{s}=4.178$~GeV. 
	The points with error bars are data, the blue solid curves are the total fits, and the black dashed curves are the fitted backgrounds. 
	The pairs of red arrows denote the signal regions.}
\label{fit:Mass-data-Ds_4180}
\end{center}
\end{figure*}

To measure the BF, we start from the following equations. 
The ST yield for each tag mode is given by
\begin{equation}
  N^{\text{ST}} = 2N_{D_{s}^{+}D_{s}^{-}}\mathcal{B}_{\text{tag}}\epsilon_{\text{tag}}^{\text{ST}}\,, \label{eq-ST}
\end{equation}
and the DT yield is given by
\begin{equation}
  N^{\text{DT}}=2N_{D_{s}^{+}D_{s}^{-}}\mathcal{B}_{\text{tag}}\mathcal{B}_{\text{sig}}\epsilon_{\text{tag,sig}}^{\text{DT}}\,,
  \label{eq-DT}
\end{equation}
where 
$N_{D_{s}^{+}D_{s}^{-}}$ is the total number of $D_{s}^{*\pm}D_{s}^{\mp}$ pairs produced in the $e^{+}e^{-}$ collisions,
$\mathcal{B}_{\text{tag}}$ and $\mathcal{B}_{\text{sig}}$ are the BFs of the tag and signal modes, respectively, $\epsilon_{\text{tag}}^{\text{ST}}$ is the ST efficiency to reconstruct the tag mode, and $\epsilon_{\text{tag,sig}}^{\text{DT}}$ is the DT efficiency to reconstruct both the tag and signal decay modes.
The total DT yield for all the tag modes $\alpha$ and all the sample groups $i$ is written as 
\begin{eqnarray}
\begin{aligned}
  \begin{array}{lr}
    N_{\text{total}}^{\text{DT}}=\Sigma_{\alpha, i}N_{\alpha,\text{sig},i}^{\text{DT}}   = \mathcal{B}_{\text{sig}}
    \Sigma_{\alpha, i}2N^{i}_{D_{s}^{+}D_{s}^{-}}\mathcal{B}_{\alpha}\epsilon_{\alpha,\text{sig}, i}^{\text{DT}}\,.
  \end{array}
  \label{eq-DTtotal}
\end{aligned}
\end{eqnarray}
Solving for $\mathcal{B}_{\text{sig}}$,
\begin{eqnarray}\begin{aligned}
	\mathcal{B}(D_s^+\to\omega\pi^+\eta) =
	\frac{N^{\rm DT}_{\rm total}}{B_{\rm sub}\sum_{\alpha,i}{N^{\rm ST}_{\alpha,i}}\epsilon^{\rm DT}_{\alpha,\rm sig,i}/\epsilon^{\rm ST}_{\alpha,i}},
	\label{abs:bf}
\end{aligned}\end{eqnarray}
where the BF $B_{\rm sub}=\mathcal{B}_{\pi^0\to\gamma\gamma}\mathcal{B}_{\eta\to\gamma\gamma}\mathcal{B}_{\omega\to\pi^+\pi^-\pi^0}$ is introduced to take into account that the signal is reconstructed through these decays. 
The ST yields $N_{\alpha,i}^{\text{ST}}$ and ST efficiencies $\epsilon_{\alpha,i}^{\text{ST}}$ are obtained from the data and inclusive MC samples, respectively.
For the  inclusive MC samples  we use the same ST selection criteria and fitting strategy as those used in data analysis, to extract the number of observed ST events by fitting the $M_{\rm tag}$ distributions, as shown in Fig.~\ref{fit:Mass-data-Ds_4180}.
The efficiencies are computed as the ratio of the number of observed ST~(DT) events and the number of generated ST~(DT) events in the inclusive MC samples.

\section{EVENT SELECTION} \label{tag_selection}
The $D_s^{\pm}$ candidates are constructed from individual $\pi^\pm$, $K^\pm$, $K^0_S$, $\pi^0$, $\eta$, and $\omega$ candidates, with the following selection criteria.

Charged tracks detected in the MDC are required to be within a polar angle ($\theta$) range of $|$cos$\theta|<0.93$, 
where $\theta$ is defined with respect to the $z$-axis, which is the symmetry axis of the MDC.
For charged tracks not originating from $K_S^0$ decays, the distance of closest approach to the interaction point~(IP) must be less than 10~cm along the $z$-axis, and less than 1~cm in the transverse plane.
Particle identification~(PID) for charged tracks combines the measurements of the ${\rm d}E/{\rm d}x$ in the MDC and the flight time in the TOF to form likelihoods $\mathcal{L}(h)~(h=K,\pi)$ for each hadron $h$ hypothesis.
Charged kaons and pions are identified by comparing the likelihoods for the two hypotheses, $\mathcal{L}(K)>\mathcal{L}(\pi)$ and $\mathcal{L}(\pi)>\mathcal{L}(K)$, respectively.

The $K_{S}^0$ candidates are selected by looping over all pairs of tracks with opposite charge, whose distances to the IP along the beam direction are less than $20$ cm. 
These two tracks are assumed to be pions without PID criteria applied.  
These tracks are required to originate from a common vertex and have a $\pi^{+}\pi^{-}$ invariant mass ($M_{\pi^{+}\pi^{-}}$) in the range $[0.487, 0.511]$ GeV$/c^{2}$, corresponding to about three times the resolution of the detector.

Photon candidates are identified using showers in the EMC. 
The deposited energy of each shower must be greater than 25~MeV in the barrel region~($|\!\cos \theta|< 0.80$) and more than 50~MeV in the end cap region~($0.86 <|\!\cos \theta|< 0.92$). 
To exclude showers originating from charged tracks, the angle subtended at the IP by the EMC shower and the position of the closest charged track extrapolated to the EMC must be greater than 10 degrees. 
The difference between the EMC time and the event start time is required to be within [0, 700]~ns to suppress electronic noise and showers unrelated to the event.

The $\pi^0$ and $\eta$ candidates are reconstructed from photon pairs with invariant masses in the ranges $[0.115, 0.150]$~GeV/$c^{2}$ and $[0.500, 0.570]$~GeV/$c^{2}$, respectively.
To improve their invariant mass resolutions, we require that at least one photon comes from the barrel region of the EMC. 
Furthermore, the $\pi^0$ candidates are constrained to the known $\pi^0$ mass~\cite{PDG} via a kinematic fit to improve their energy and momentum resolution.

In order to remove soft pions from $D^{*+}$ decays, both charged and neutral pion candidates are subjected to an additional  momentum cut, $p(\pi) > 100$ MeV$/c$.

\begin{table}[htbp]
 \centering
	\caption{Requirements on $M_{\rm tag}$ for various tag modes.}
     \begin{tabular}{lc}
        \hline
        Tag mode                                     & Mass window (GeV/$c^{2}$) \\
        \hline
        $D_{s}^{-} \to K_{S}^{0}K^{-}$               & [1.948, 1.991]            \\
        $D_{s}^{-} \to K^{+}K^{-}\pi^{-}$            & [1.950, 1.986]            \\
        $D_{s}^{-} \to K^{+}K^{-}\pi^{-}\pi^{0}$     & [1.947, 1.982]            \\
        \hline
      \end{tabular}
	\label{tab:tag-cut}
\end{table}
Three tag modes are used to reconstruct the ST $D^-_s$ candidate, and the selections on the invariant mass $(M_{\rm tag})$ for each tag mode are listed in Table~\ref{tab:tag-cut}.
The tag mode $D_{s}^{-}\to K^-K^+\pi^-$ is mainly from $\phi \pi^-$ or $\bar K^{*0}(892)K^-$.
Therefore, we require that $M_{K^-K^+} <$ 1.05 GeV/$c^2$ or $|M_{K^+\pi^-}-0.892| <$ 0.070 GeV/$c^2$, where $M_{K^-K^+}$ and $M_{K^+\pi^-}$ are the invariant masses of $K^-K^+$ and $K^+\pi^-$, respectively.
The tag mode $D_{s}^{-}\to K^-K^+\pi^-\pi^0$ is mainly from $\phi \rho^-$.
Therefore, we require $M_{K^-K^+} <$ 1.05 GeV/$c^2$ and $|M_{\pi^-\pi^0} -0.775|<$ 0.150 GeV/$c^2$, where $M_{K^-K^+}$ and $M_{\pi^-\pi^0}$ are the invariant masses of $K^-K^+$ and $\pi^-\pi^0$.
To further suppress backgrounds from non-$D^{*\pm}_s D^\mp_s$ processes, a selection on the variable $M_{\rm rec}$ is applied, defined as
\begin{eqnarray}
\begin{aligned}
  \begin{array}{lr}
		M_{\rm rec} = \sqrt{\left(p_{\rm cm} - p_{D_{s}^-}\right)^{2}}, \label{eq:mrec}
	\end{array}
\end{aligned}
\end{eqnarray}
where $p_{\rm cm}$ is the four-momentum of the $e^+e^-$ center-of-mass system, 
$p_{D_{s}^-}$ is the four-momentum of the $D_{s}^{-}$ candidate in the $e^+e^-$ center-of-mass frame.
The mass windows of $M_{\rm rec}$ for $D^-_s$ candidates at various center-of-mass energies are listed in Table~\ref{energe}.

\begin{table*}[htbp]
  \caption{The ST yields~($N^{\rm ST}$) and ST efficiencies~($\epsilon^{\rm ST}$) for the data samples collected at $\sqrt{s} =$ (I) 4.128 and 4.157~GeV, (II) 4.178~GeV, (III) 4.189-4.219~GeV, (IV) 4.226~GeV. The uncertainties are statistical.}
  \centering
  \begin{tabular}{l r@{$\pm$}l r@{$\pm$}l r@{$\pm$}l r@{$\pm$}l r@{$\pm$}l r@{$\pm$}l r@{$\pm$}l r@{$\pm$}l}
  \hline
    Data sample &\multicolumn{4}{c}{(I)}  &\multicolumn{4}{c}{(II)}  &\multicolumn{4}{c}{(III)}  &\multicolumn{4}{c}{(IV)}\\
  \hline
    Tag mode   & \multicolumn{2}{c}{$N^{\rm ST}$} &\multicolumn{2}{c}{$\epsilon^{\rm ST}$ $(\%)$} 
    &\multicolumn{2}{c}{$N^{\rm ST}$}  &\multicolumn{2}{c}{$\epsilon^{\rm ST}~(\%)$}
    &\multicolumn{2}{c}{$N^{\rm ST}$}  &\multicolumn{2}{c}{$\epsilon^{\rm ST}~(\%)$}
    &\multicolumn{2}{c}{$N^{\rm ST}$}  &\multicolumn{2}{c}{$\epsilon^{\rm ST}~(\%)$}\\
  \hline
    $D^-_s\to K^0_SK^-$          &6728  &144 &47.66 &0.17 &31957  &314 &47.40 &0.08 &19960 &270 &47.20 &0.09  &6836  &163  &47.92 &0.18\\
    $D^-_s\to K^+K^-\pi^-$       &23443 &202 &33.67 &0.05 &114890 &447 &32.88 &0.02 &72827 &369 &32.69 &0.03  &24862 &227  &33.03  &0.05 \\
    $D^-_s\to K^+K^-\pi^-\pi^0$  &2414  &102 &3.43&0.02 &13304  &269 &3.59 &0.01 &8586  &227 &3.70  &0.02  &3172 &146   &3.84  &0.03\\
    \hline
  \end{tabular}
  \label{ST-eff}
\end{table*}
	For multiple ST candidates, the candidate with $M_{\rm rec}$ closest to the known mass of $D_s^{*+}$~\cite{PDG} is chosen as the signal candidate. 
Only $0.16\%$ ($0.17\%$) of events in data (simulated samples) contain multiple $D^+_s$ candidates, and the multiplicity distribution of the ST candidates shows consistency between data and simulated samples.
  Table~\ref{ST-eff} lists the yields and efficiencies for various tag modes, in which the yields at (I) and (III) are fitted with combined data sets.
	We use the same ST selection criteria and fitting strategy as those used in data analysis to analyse the inclusive MC samples, and extract the number of observed ST events from fitting the $M_{\rm tag}$ distribution. 
	The ST efficiency is computed as the ratio of the number of observed ST events and the number of generated ST events in the inclusive MC samples. 
As an example, the fits to the accepted ST candidates from the data sample at $\sqrt s = 4.178$~GeV are shown in Fig.~\ref{fit:Mass-data-Ds_4180}.
In the fits, the signal is modeled by an MC-simulated shape convolved with a Gaussian function taking into account the data-MC resolution difference. 
The background is described by a second-order Chebyshev polynomial. 
For the tag mode $D_{s}^{-} \to K_{S}^{0} K^-$, the peaking background from $D^{-} \to K_{S}^{0} \pi^-$ is included in the fit, with a shape taken from the inclusive MC samples and its yield floating.  
\begin{figure}[htbp]
          \centering
          \includegraphics[width=7.5cm]{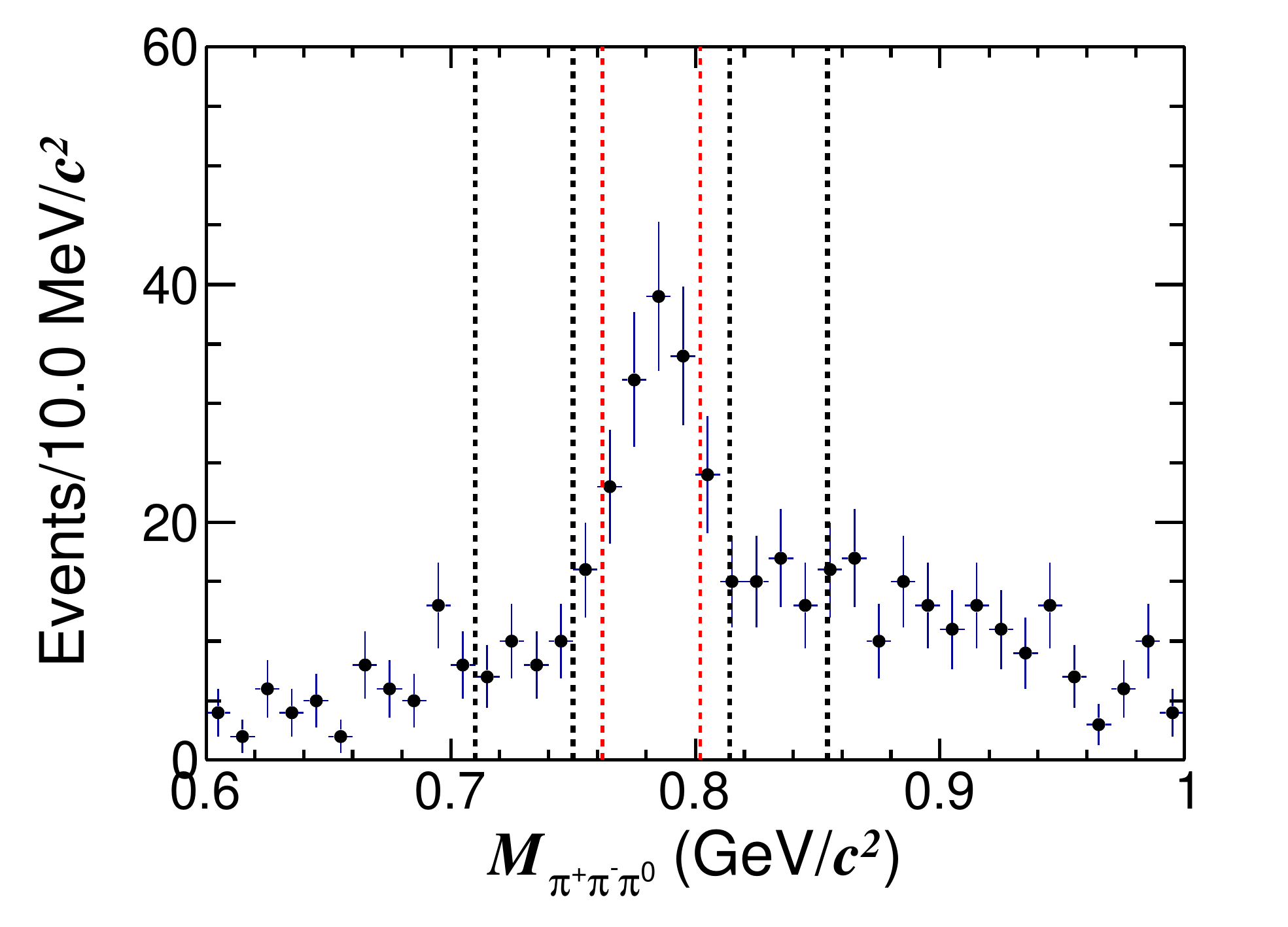}
		\caption{Invariant mass distribution for selected $\omega\to\pi^+\pi^-\pi^0$ candidates at $\sqrt{s}=4.128-4.226$ GeV. 
		The points with error bars are data.
		The red vertical dashed lines indicate the $\omega$ signal region, and the pairs of black dashed lines (left and right of the signal peak) indicate the $\omega$ sideband regions.}
    \label{omega}
\end{figure}

Once a tag mode is identified, we search for the signal decay $D_{s}^{+} \to \omega\pi^+\eta$ among the recoiling particles, with the $\omega$ and $\eta$ candidates reconstructed in the $\pi^+\pi^-\pi^0$ and $\gamma\gamma$ final states, respectively. 
The $\pi^+\pi^-\pi^0$ combination with minimum $|M_{\pi^+\pi^-\pi^0}-m_{\omega}|$ is used to reconstruct the $\omega$ candidates, where $m_{\omega}$ is the known mass of the $\omega$ taken from the PDG~\cite{PDG}.
The invariant mass distribution of $\pi^+\pi^-\pi^0$ for data is shown in Fig.~\ref{omega}, in which a clear omega peak is observed. 
There is no obvious background from $D_s^+ \to \eta \pi^+\pi^- \pi^+ \pi^0$ in the $\omega$ sideband regions ([0.71,0.75] and [0.814,0.854] GeV/$c^2$).
Therefore we require the $\pi^+\pi^-\pi^0$ invariant mass to be in the omega signal range [0.762, 0.802] GeV/$c^2$ for further analysis.
In the case of multiple candidates, the DT candidate with the average mass, $(M_{\rm sig}+M_{\rm tag})/2$, closest to the known $D_s^+$ mass is retained. 
The $M_{\rm sig}$ is defined as the invariant-mass of the accepted signal $D^+_s$ candidates.

To suppress the background from the decay $D^+_s\to\pi^+\pi^0\eta^{\prime}$ with $\eta^{\prime}\to\pi^+\pi^-\eta$, we reject candidates with the $\pi^+\pi^-\eta$ invariant mass less than 1 GeV/c$^2$. 
A kinematic fit is performed under the hypothesis $e^+e^- \to D_s^{*\pm} D_s^\mp \to \gamma D_s^+ D_s^-$, with $D_s^-$ decaying to one of the tag modes and $D_s^+$ decaying to the signal mode. The combination with the minimum $\chi^2$ assuming a $D_s^{*+}$ meson decays to $D_s^+ \gamma$ or a $D_s^{*-}$ meson decays to $D_s^- \gamma$ is chosen to reconstruct the transition photon of $D_s^{*\pm} \to \gamma D_s^\pm$.  
To suppress the non-$D^{*+}_sD_s^-$ background, we require the missing energy~$(E_{\rm {missing}} = E_{\rm cm} - E_{\rm tag} - E_{\rm sig} - E_{\gamma})$ to be less than 0.08 GeV, 
where $E_{\rm tag}$, $E_{\rm sig}$, and $E_{\gamma}$ are the energies of $D_{s}^{-}$, $D_{s}^{+}$, 
and transition photon of $D_s^{*\pm} \to \gamma D_s^\pm$ in the $e^+e^-$ center-of-mass frame, respectively.

\section{BRANCHING FRACTION MEASUREMENT}
\begin{figure*}[htbp]
          \centering
          \includegraphics[width=5.9cm]{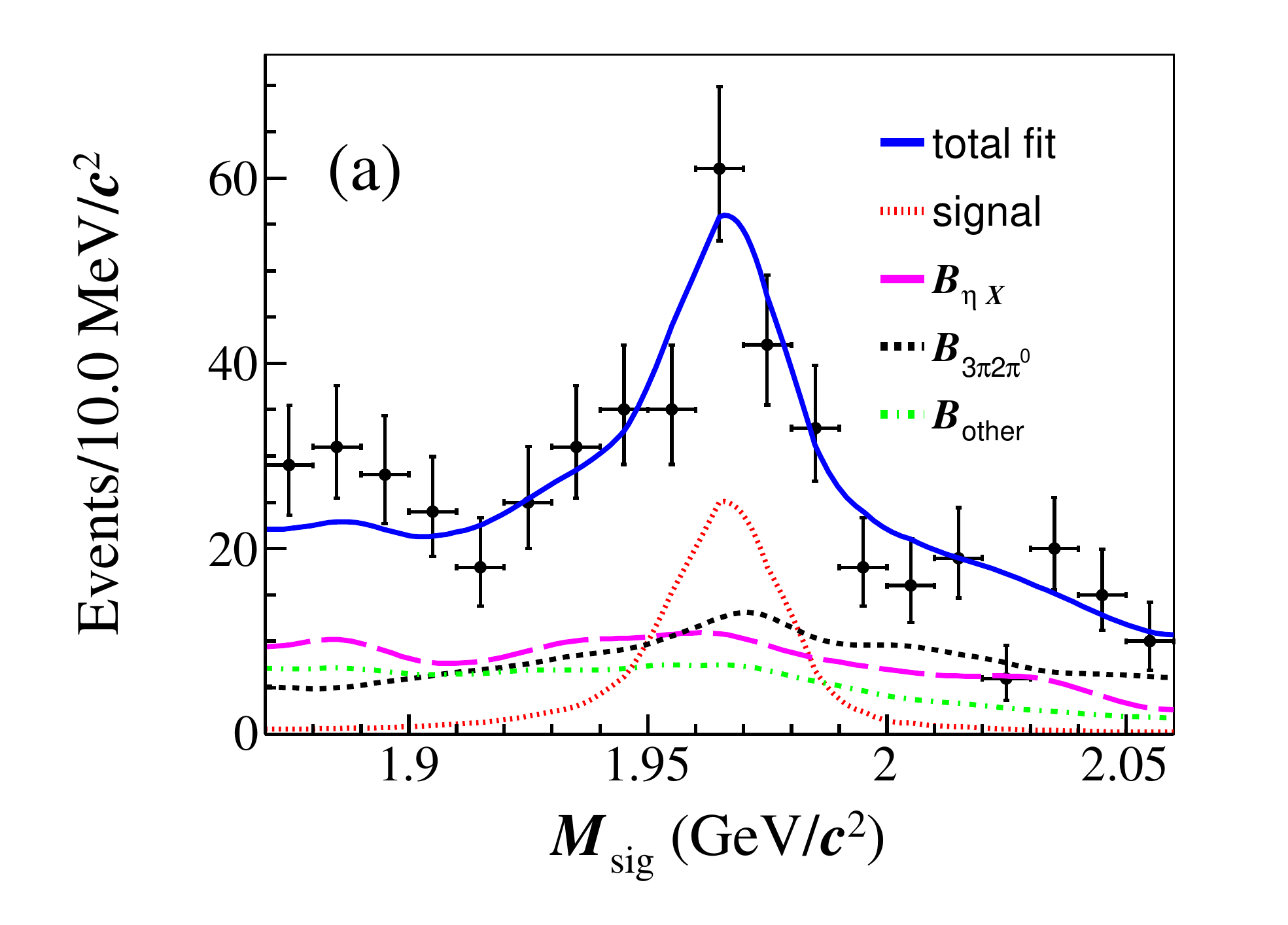}
          \includegraphics[width=5.9cm]{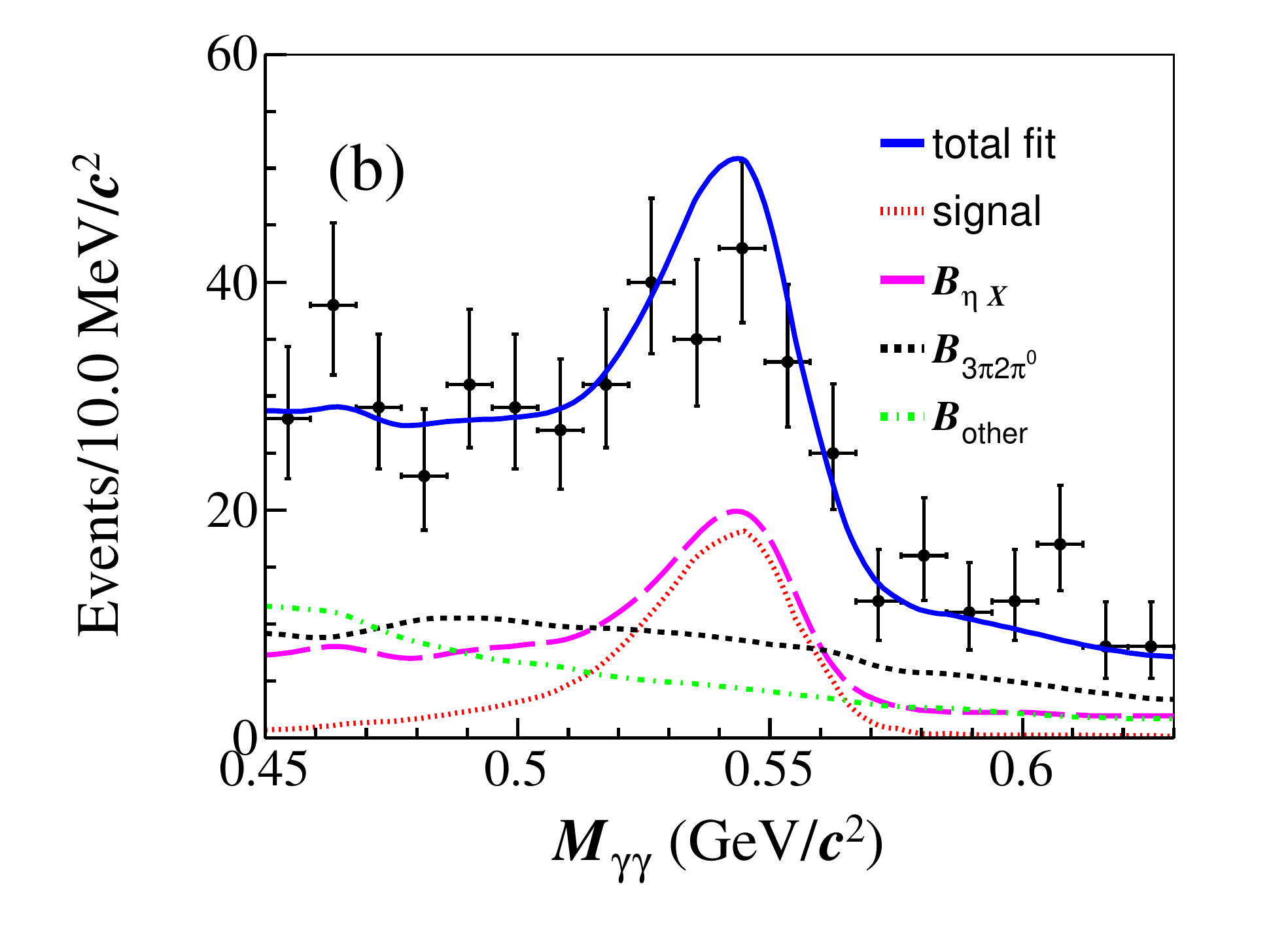}
					\includegraphics[width=5.9cm]{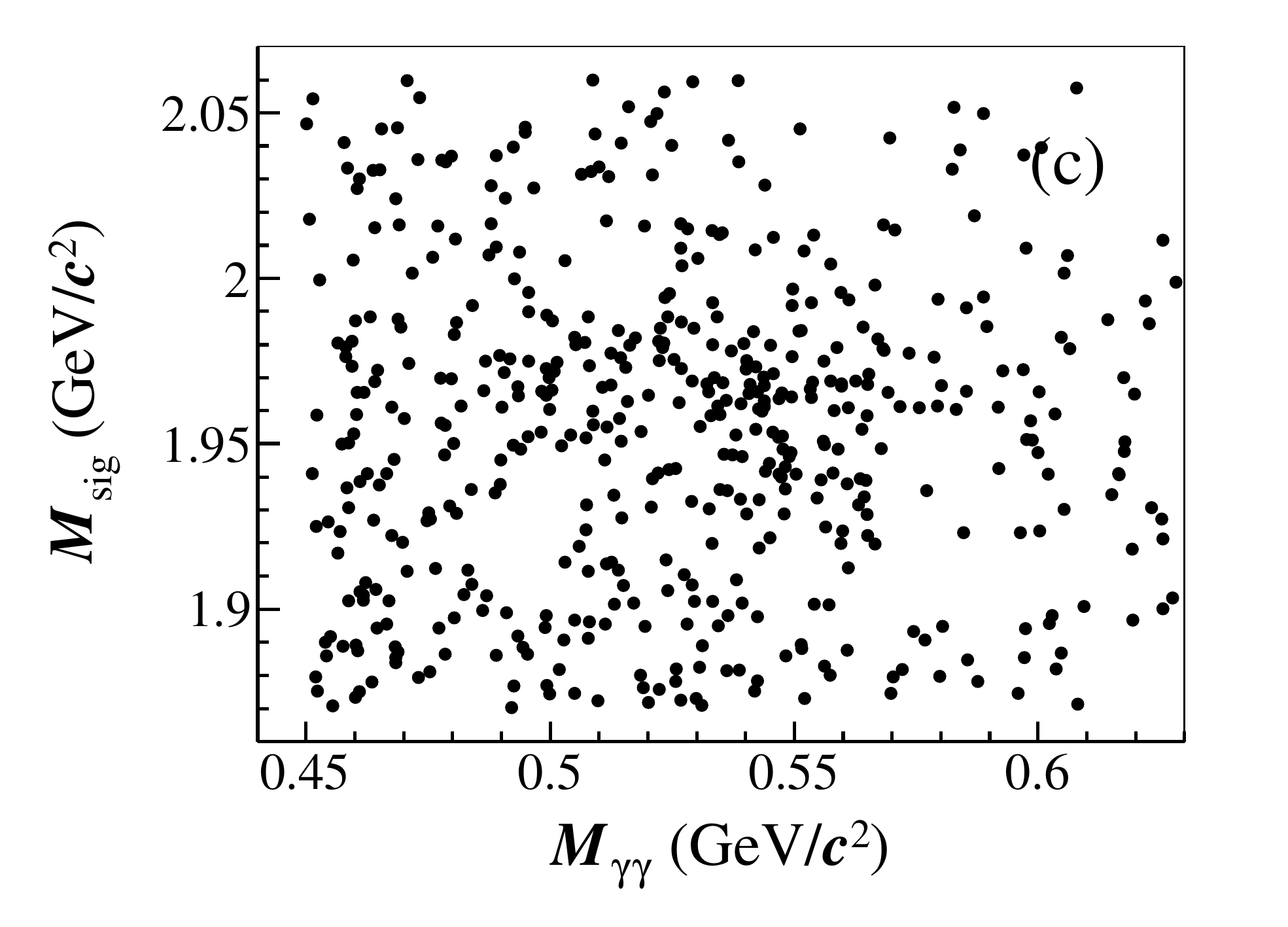}
    \caption{
		Distributions of (a)~$M_{\rm sig}$, (b)~$M_{\gamma\gamma}$, and (c)~$M_{\rm sig}$ versus $M_{\gamma\gamma}$ at $\sqrt{s} = 4.128 - 4.226$ GeV.
	The points with error bars are data, the solid blue curves are the total fit; other curves show the isolated signal and the individual background contributions from the three labeled sources.}
    \label{2dfit}
\end{figure*}

The distributions of the signal $D_s^+$ candidate invariant mass~($M_{\rm sig}$), the $\gamma\gamma$ invariant mass~($M_{\gamma\gamma}$), and their two-dimensional scatter plot are shown in Figs.~\ref{2dfit}~(a), (b), and (c), respectively.
It is possible to distinguish three kinds of background events. 
The events which contain $\eta$ in final states are called $B_{\eta X}$;
they are peaking in $M_{\gamma\gamma}$ but flat in $M_{\rm sig}$. 
The events from $D^{+}_{s}\to \pi^0\pi^0\pi^+\pi^+\pi^-$, similar to the signal with an incorrectly reconstructed $\eta$, are called $B_{3\pi2\pi^0}$;
they are peaking in $M_{\rm sig}$ and flat in $M_{\gamma\gamma}$. 
The remaining background events are called $B_{\rm other}$.

 The signal yield is extracted from a two-dimensional (2D) unbinned maximum-likelihood fit to the distributions of $M_{\rm sig}$ and $M_{\gamma\gamma}$ at all energy points.
The signal shape is described by an MC-simulated 2D probability density function (PDF) convolved with a Gaussian function, whose parameters are derived from the corresponding $M_{\rm sig}$ and $M_{\gamma\gamma}$ fits,
to consider resolution the difference between data and MC simulation.
The shapes for various background components are modeled from the inclusive MC samples, where the model for $B_{3\pi2\pi^0}$ is obtained from the amplitude analyses of the decays $D^{+}_{s}\to \pi^0\pi^+\eta~(\eta\to\pi^0\pi^+\pi^-$)~\cite{BESIII:2019jjr} and $D^{+}_{s}\to \pi^0\pi^0\pi^+\pi^+\pi^-$. 
The yield of $B_{3\pi2\pi^0}$ is fixed to $152.9$ according to the measured BFs, while the yields of $B_{\eta X}$ and $B_{\rm other}$ are floating parameters in the fit.

\begin{table*}[htbp]
	\caption{DT efficiencies~($\epsilon^{\rm DT}$) for the data samples taken at (I)$\sqrt{s}=4.128$ and $4.157$ GeV, (II)$\sqrt{s}=4.178$ GeV, (III)$\sqrt{s}=4.189-4.219$ GeV, and (IV)$\sqrt{s}=4.226$ GeV. 
	The BFs of the sub-particle~($K^0_S,\pi^0$) decays are not included.}
  \centering
  \begin{tabular}{lcccc}
  \hline
    Tag mode   &(I)$\epsilon_{\rm DT}$ $(\%)$ &(II)$\epsilon_{\rm DT}$ $(\%)$    &(III)$\epsilon_{\rm DT}$ $(\%)$ &(IV)$\epsilon_{\rm DT}$ $(\%)$\\
  \hline
    $D^-_s\to K^0_SK^-$           &$6.02\pm0.53$ &$5.96\pm0.23$ &$5.67\pm0.28$ &$5.98\pm0.49$\\
		$D^-_s\to K^+K^-\pi^-$        &$4.25\pm0.20$ &$4.13\pm0.08$ &$4.06\pm0.11$ &$3.83\pm0.17$\\
		$D^-_s\to K^+K^-\pi^-\pi^0$   &$0.38\pm0.06$ &$0.45\pm0.03$ &$0.43\pm0.03$ &$0.45\pm0.06$\\
    \hline
  \end{tabular}
  \label{abs:dteff}
\end{table*}
From the 2D fit, we obtain $78\pm16$ $D^+_s\to\omega\pi^+\eta$ signal events with a statistical significance of 7.6$\sigma$.
Along with the DT, the corresponding efficiencies are determined and listed in Table~\ref{abs:dteff}. 
We use the same DT selection criteria as those used in data analysis to analyse the inclusive MC samples, and extract the number of obtained DT events from counting the signal events.
  The DT efficiency is computed as the ratio of the number of obtained DT events and the number of generated DT events in the inclusive MC samples, the yields at $\sqrt{s}=4.128$ and $4.157$ GeV, and $\sqrt{s}=4.189-4.219$ GeV are obtained with combined data sets.
The statistical significance is evaluated using $\sqrt{-2{\rm ln}(\mathcal {L}_0/\mathcal {L}_{\rm max})}$, where $\mathcal {L}_{\rm max}$ is the maximum likelihood of the nominal fit and $\mathcal {L}_0$ is the likelihood of the fit excluding the signal PDF.
With Eq.~(\ref{abs:bf}), the BF is measured to be $\mathcal{B}(D^+_s\to\omega\pi^+\eta) = (0.54\pm0.12)\%$, where the uncertainty is statistical. 

\begin{figure*}[htbp]
          \centering
          \includegraphics[width=7.5cm]{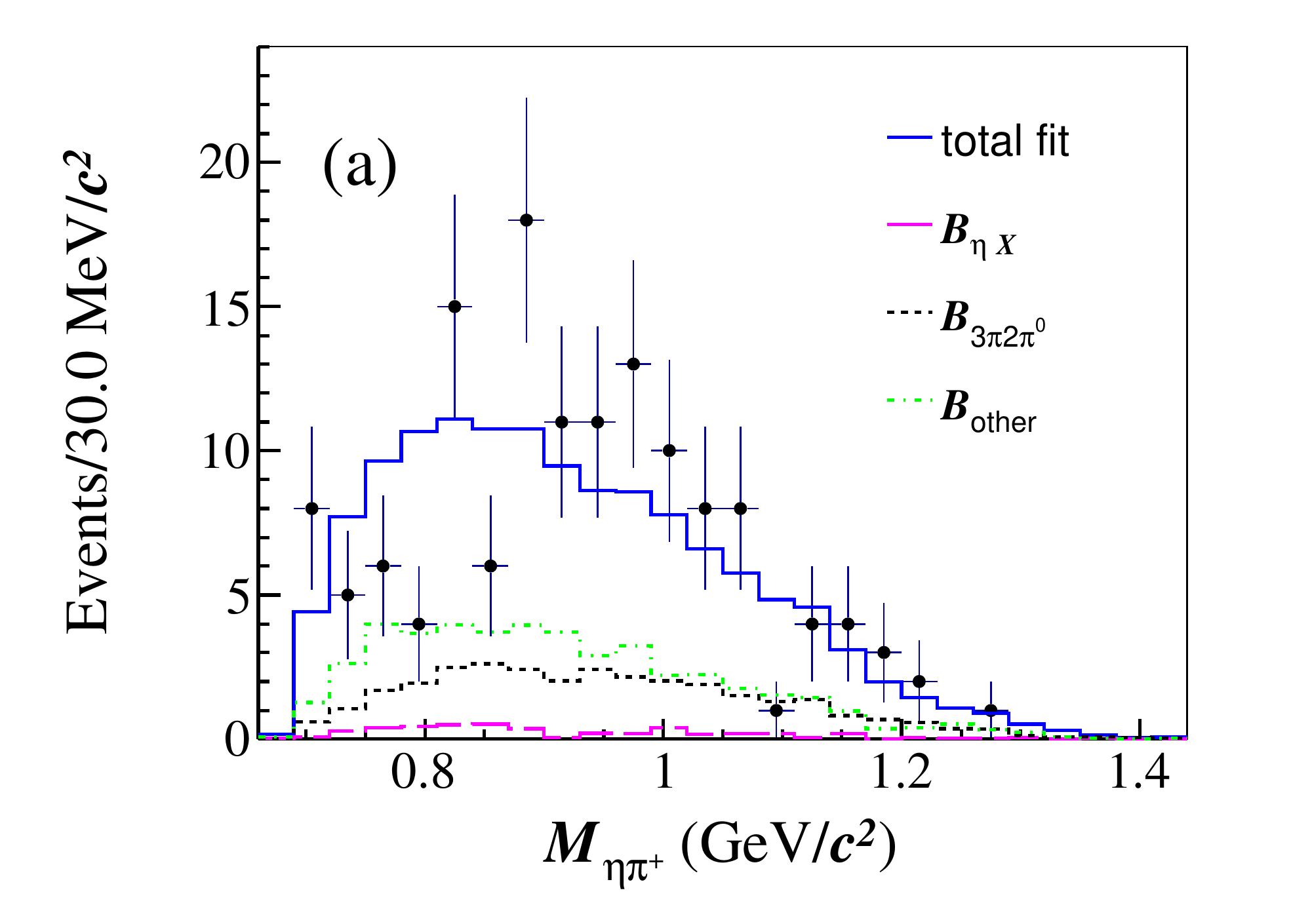}
          \includegraphics[width=7.5cm]{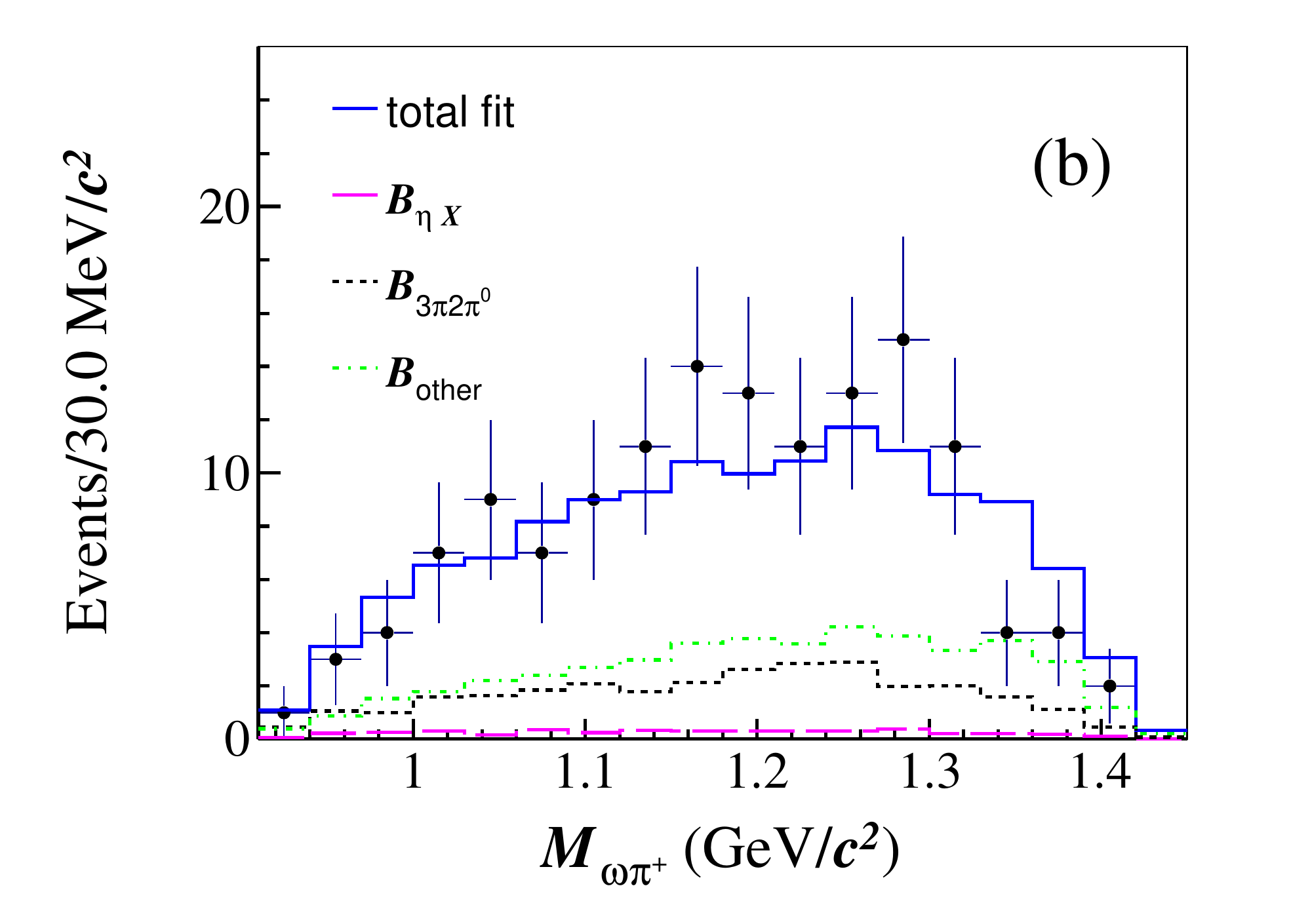}
          \includegraphics[width=7.5cm]{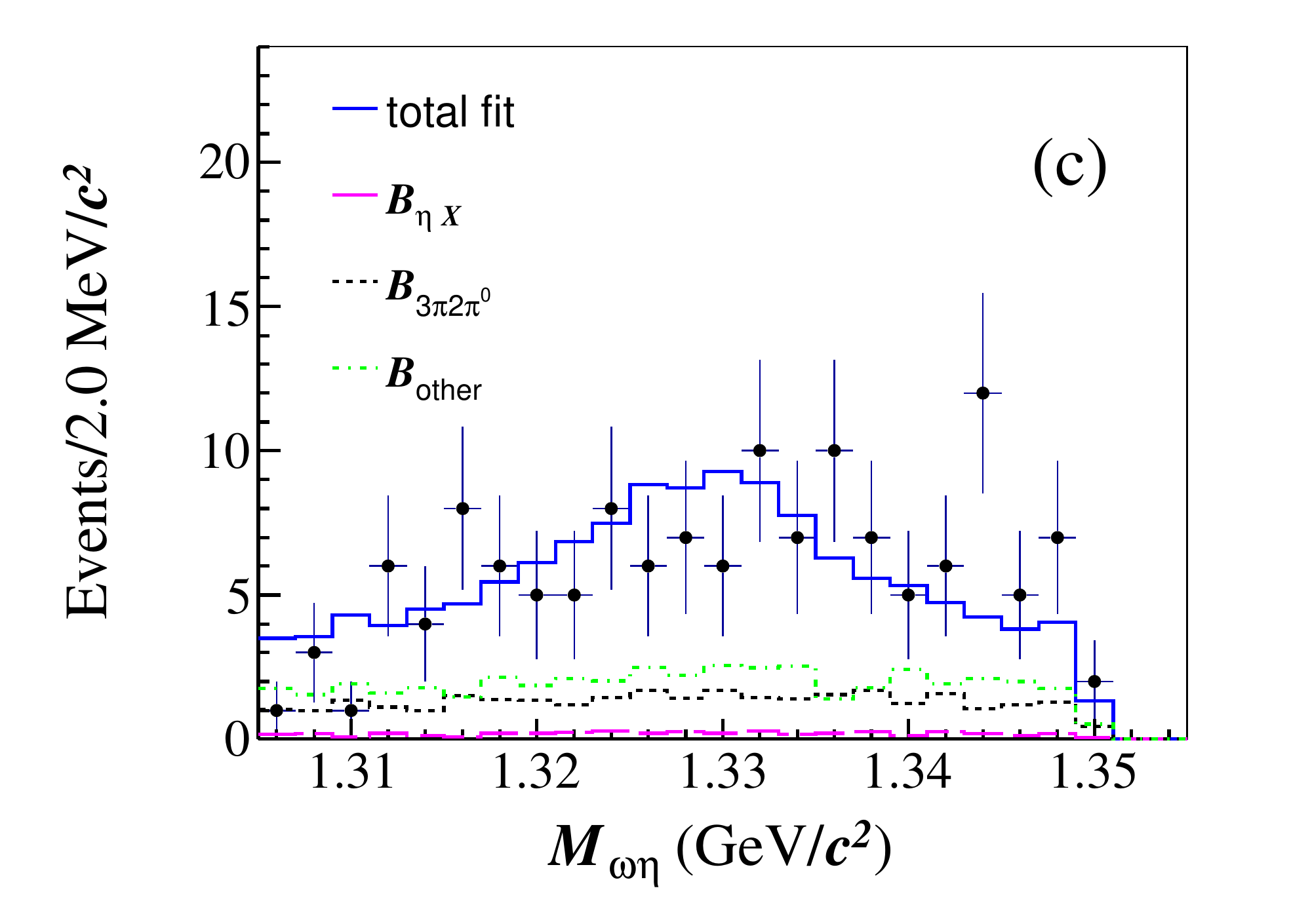}
          \includegraphics[width=7.4cm]{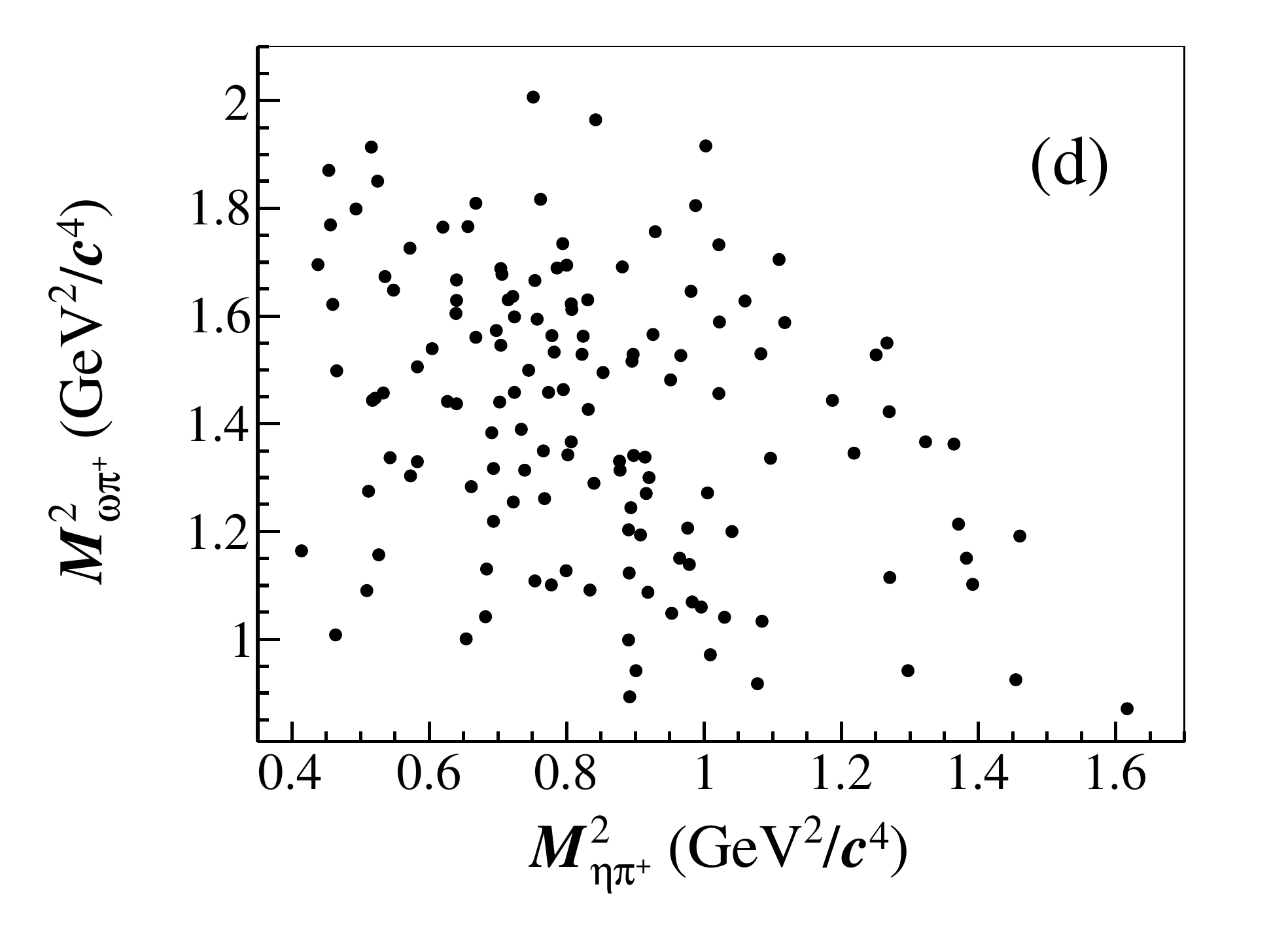}
	\caption{Mass projections onto (a)~$M_{\eta\pi^+}$, (b)~$M_{\omega\pi^+}$, (c)~$M_{\omega\eta}$ and the distribution of (d)~$M^2_{\omega\pi^+}$ versus $M^2_{\eta\pi^+}$ from the combined data sample at $\sqrt{s}= 4.128-4.226$ GeV.
		In (a), (b) and (c), the points with error bars are data and the solid blue curves are the total fit projections; other curves show the individual background contributions from the three labeled sources.}
          \label{project}
\end{figure*}

The resulting projections on $M_{\eta\pi^+}$, $M_{\omega\pi^+}$ and $M_{\omega\eta}$, as well as the distribution of $M^2_{\omega\pi^+}$ versus $M^2_{\eta\pi^+}$ for the data in the signal region ($M_{\rm sig}\in (1.94,1.99)$ and $M_{\rm \gamma\gamma}\in (0.49,0.57)$~GeV/$c^2$), are shown in Fig.~\ref{project}.
Due to the limited data sample size, no sub-resonances such as $a^+_0(980)$ or $b_1(1235)^+$ can be identified.

\section{SYSTEMATIC UNCERTAINTY}
The different systematic uncertainties in the BF measurement are discussed below;
most systematic uncertainties related to the efficiency for reconstructing the tag side cancel due to the DT technique.

\begin{itemize}
\item ST yield. 
The total ST yield of the three tag modes is $328979\pm2880$,
		resulting in statistical uncertainty $\sqrt{(2880^{2}-328979)}/328979 = 0.3\%$.
Here, 
		we only consider the statistical fluctuation related to the background of the tag side which is not correlated with the DT sample directly; hence,  $0.3\%$ is assigned as a systematic uncertainty.

\item $\pi^\pm$ tracking and PID efficiencies. 
The systematic uncertainties in the tracking and PID efficiencies per charged pion are assigned to be 1.0$\%$ and 1.0$\%$, respectively, estimated by the control sample $e^+e^-\to K^-K^+\pi^-\pi^+$. 

\item 2D fit. 
	To estimate the systematic uncertainty related with the signal shape, we observe the change in BF when varying the mean and resolution of the convolving Gaussian function by their corresponding uncertainties.
We take the variation of the BF, 1.7$\%$, as the systematic uncertainty. 
The systematic uncertainty due to the MC-simulated background shape is studied by varying the relative fractions of the backgrounds from $q\bar q$ and non-$D_s^{*+}D_s^-$ open charm by $\pm 30\%$~\cite{BESIII:2021xox}, the statistical uncertainty of their cross sections. 
The largest change of the BF, 0.8$\%$, is taken as the systematic uncertainty. 
The yield of $D^{+}_{s}\to \pi^0\pi^0\pi^+\pi^+\pi^-$ is varied by its uncertainty and the change of the BF of 1.0$\%$ is assigned as the systematic uncertainty. 
The systematic uncertainty due to the signal and background on smooth number is estimated by varying parameter that describing the smoothness of the shape from 1.5 to 2 or 1.
The change of the BF, 1.3$\%$, is assigned as the systematic uncertainty. 
The total systematic uncertainty in the 2D fit is obtained to be 2.5$\%$, by adding systematic uncertainties from all above sources in quadrature.

\item MC statistics. 
An uncertainty of 2.3\% due to the limited MC statistics is obtained by $\sqrt{\begin{matrix} \sum_{i} (f_{i}\frac{\delta_{\epsilon_{i}}}{\epsilon_{i}}\end{matrix}})^2$,
where $f_{i}$ is the tag yield fraction, and $\epsilon_{i}$ and $\delta_{\epsilon_{i}}$ are the signal efficiency and the corresponding uncertainty of tag mode $i$, respectively.

\item $\pi^0$ and $\eta$ reconstruction efficiencies. 
The systematic uncertainty of the $\pi^0$ reconstruction efficiency is estimated to be 2.0$\%$ with a control sample of $e^+e^-\to K^-K^+\pi^-\pi^+\pi^0$. 
The systematic uncertainty of the $\eta$ reconstruction efficiency is taken to be 2.0$\%$, based on the $\pi^0$ uncertainty. 
		The total systematic uncertainty of 4.0\% due to $\pi^0$ and $\eta$ reconstructions is obtained by adding them linearly.

\item Quoted BFs. 
	The uncertainties on the quoted BFs of $\pi^0\to\gamma\gamma$, $\eta\to\gamma\gamma$ and $\omega\to\pi^+\pi^-\pi^0$ are $0.03\%$ (negligible), $0.5\%$ and $0.8\%$, respectively~\cite{PDG}.

\item $\omega$ mass window. 
The systematic uncertainty due to the $\omega$ mass window is studied by using a control sample of the decay $D^0 \to K^-\pi^+\omega$.
		The difference in the efficiencies of the $\omega$ mass window between data and MC simulation, $1.2\%$, is taken as the corresponding systematic uncertainty.

\item MC model. 
	To estimate the systematic uncertainty related with the MC model, we separately add $D_s^+\to\omega a_0(980)^+$ or $D_s^+\to\eta b_1(1235)^+$ signal MC events to inclusive MC samples to improve data-MC consistency (based on 
agreement of mass projections). The larger change of the signal efficiency, 1.0$\%$, is taken as the related systematic uncertainty.
\end{itemize}

All of the systematic uncertainties are summarized in Table~\ref{BF-Sys}.
		Adding them in quadrature gives a total systematic uncertainty in the BF measurement of 6.6\%.
\begin{table}[htbp]
  \caption{Relative systematic uncertainties of the BF measurement.}
	\centering
    \begin{tabular}{lccc}
    \hline
    Source                                  &Uncertainty (\%)\\
    \hline
		ST yield                                &0.3\\
    Tracking                                &3.0\\
    PID                                     &3.0\\
    2D fit                                  &2.5\\
    MC statistics                           &2.3\\
    $\pi^0$ and $\eta$ reconstruction       &4.0\\
    $\mathcal{B}(\eta\to\gamma\gamma)$      &0.5\\
    $\mathcal{B}(\omega\to\pi^+\pi^-\pi^0)$ &0.8\\
    $\omega$ mass                         &1.2\\
    MC model                                  &1.0\\
    \hline
		Total                                   &7.0\\
    \hline
    \end{tabular}
  \label{BF-Sys}
\end{table}

\section{Summary}
In summary, using 7.33~fb$^{-1}$ of $e^+e^-$ collision data collected with the BESIII detector between $\sqrt{s} = 4.128$ and $4.226$ GeV, we have reported the first observation of $D^+_s\to\omega\pi^+\eta$ with a statistical significance of $7.6\sigma$. 
The absolute BF of this decay is measured to be ${\cal B}(D^+_s\to \omega\pi^+\eta) = (0.54\pm0.12\pm0.04)\%$, where the first uncertainty is statistical and the second is systematic. 
The BF measured in this work is consistent with that of the CLEO collaboration~\cite{CLEO:2009nsf}, but the precision is improved by a factor of 2.7.

Our result offers an important input for estimating the $D^+_s\to\pi^+\pi^+\pi^- X$ background contribution in tests of the LFU with semileptonic $B$ decays.
Larger statistics data to be taken in the future~\cite{Ablikim:2019hff} will help to search for potential intermediate processes $D_s^+ \to \omega a_0(980)^{+}$ and $D_s^+ \to \eta b_1(1235)^+$. 

\section*{acknowledgments}
The BESIII collaboration thanks the staff of BEPCII and the IHEP computing center for their strong support. This work is supported in part by National Key R\&D Program of China under Contracts Nos. 2020YFA0406400, 2020YFA0406300; National Natural Science Foundation of China (NSFC) under Contracts Nos. 11635010, 11735014, 11835012, 11935015, 11935016, 11935018, 11961141012, 12022510, 12025502, 12035009, 12035013, 12192260, 12192261, 12192262, 12192263, 12192264, 12192265; the Chinese Academy of Sciences (CAS) Large-Scale Scientific Facility Program; Joint Large-Scale Scientific Facility Funds of the NSFC and CAS under Contract No. U1832207, U1932108; the CAS Center for Excellence in Particle Physics (CCEPP); 100 Talents Program of CAS; The Institute of Nuclear and Particle Physics (INPAC) and Shanghai Key Laboratory for Particle Physics and Cosmology; ERC under Contract No. 758462; European Union's Horizon 2020 research and innovation programme under Marie Sklodowska-Curie grant agreement under Contract No. 894790; German Research Foundation DFG under Contracts Nos. 443159800, 455635585, Collaborative Research Center CRC 1044, FOR5327, GRK 2149; Istituto Nazionale di Fisica Nucleare, Italy; Ministry of Development of Turkey under Contract No. DPT2006K-120470; National Science and Technology fund; National Science Research and Innovation Fund (NSRF) via the Program Management Unit for Human Resources \& Institutional Development, Research and Innovation under Contract No. B16F640076; Olle Engkvist Foundation under Contract No. 200-0605; STFC (United Kingdom); Suranaree University of Technology (SUT), Thailand Science Research and Innovation (TSRI), and National Science Research and Innovation Fund (NSRF) under Contract No. 160355; The Royal Society, UK under Contracts Nos. DH140054, DH160214; The Swedish Research Council; U. S. Department of Energy under Contract No. DE-FG02-05ER41374.


\begin{thebibliography}{0}%
\makeatletter
\providecommand \@ifxundefined [1]{%
 \@ifx{#1\undefined}
}%
\providecommand \@ifnum [1]{%
 \ifnum #1\expandafter \@firstoftwo
 \else \expandafter \@secondoftwo
 \fi
}%
\providecommand \@ifx [1]{%
 \ifx #1\expandafter \@firstoftwo
 \else \expandafter \@secondoftwo
 \fi
}%
\providecommand \natexlab [1]{#1}%
\providecommand \enquote  [1]{``#1''}%
\providecommand \bibnamefont  [1]{#1}%
\providecommand \bibfnamefont [1]{#1}%
\providecommand \citenamefont [1]{#1}%
\providecommand \href@noop [0]{\@secondoftwo}%
\providecommand \href [0]{\begingroup \@sanitize@url \@href}%
\providecommand \@href[1]{\@@startlink{#1}\@@href}%
\providecommand \@@href[1]{\endgroup#1\@@endlink}%
\providecommand \@sanitize@url [0]{\catcode `\\12\catcode `\$12\catcode
  `\&12\catcode `\#12\catcode `\^12\catcode `\_12\catcode `\%12\relax}%
\providecommand \@@startlink[1]{}%
\providecommand \@@endlink[0]{}%
\providecommand \url  [0]{\begingroup\@sanitize@url \@url }%
\providecommand \@url [1]{\endgroup\@href {#1}{\urlprefix }}%
\providecommand \urlprefix  [0]{URL }%
\providecommand \Eprint [0]{\href }%
\providecommand \doibase [0]{http://dx.doi.org/}%
\providecommand \selectlanguage [0]{\@gobble}%
\providecommand \bibinfo  [0]{\@secondoftwo}%
\providecommand \bibfield  [0]{\@secondoftwo}%
\providecommand \translation [1]{[#1]}%
\providecommand \BibitemOpen [0]{}%
\providecommand \bibitemStop [0]{}%
\providecommand \bibitemNoStop [0]{.\EOS\space}%
\providecommand \EOS [0]{\spacefactor3000\relax}%
\providecommand \BibitemShut  [1]{\csname bibitem#1\endcsname}%
\let\auto@bib@innerbib\@empty
\end{thebibliography}%


\begin{thebibliography}{999}
\bibitem{HFLAV:2019otj}
Y.~S.~Amhis \textit{et al.} (HFLAV Collaboration), \href{https://link.springer.com/article/10.1140/epjc/s10052-020-8156-7}{Eur. Phys. J. C \textbf{81}, 226 (2021)};
		Updated results available at \href{https://hflav-eos.web.cern.ch/hflav-eos/semi/spring21/html/RDsDsstar/RDRDs.html} {https://hflav-eos.web.cern.ch/hflav-eos/semi/spring21/html/RDsDsstar/RDRDs.html}.

\bibitem{LHCb:2017smo}
	R.~Aaij \textit{et al.} (LHCb Collaboration), \href{https://journals.aps.org/prl/abstract/10.1103/PhysRevLett.120.171802} {Phys. Rev. Lett. \textbf{120}, 171802 (2018)}.

\bibitem{LHCb:2017rln}
	R.~Aaij \textit{et al.} (LHCb Collaboration), \href{https://journals.aps.org/prd/abstract/10.1103/PhysRevD.97.072013}{Phys. Rev. D \textbf{97}, 072013 (2018)}.

\bibitem{Li:2021iwf}
	H.~B.~Li and X.~R.~Lyu, \href{https://academic.oup.com/nsr/article/8/11/nwab181/6381732?login=true}{Natl. Sci. Rev. \textbf{8}, nwab181 (2021)}.

\bibitem{PDG}
    R.~L. Workman {\it  et al.} (Particle Data Group), {Prog. Theor. Exp. Phys. \textbf{2022}, 083C01 (2022)}.

\bibitem{Gronau:2009vt}
    M.~Gronau and J.~L.~Rosner, \href{https://journals.aps.org/prd/abstract/10.1103/PhysRevD.79.074022}{Phys. Rev. D \textbf{79}, 074022 (2009)}.

\bibitem{CLEO:2009nsf}
	J.~Y.~Ge \textit{et al.} (CLEO Collaboration), \href{https://journals.aps.org/prd/abstract/10.1103/PhysRevD.80.051102}{Phys. Rev. D \textbf{80}, 051102 (2009)}.

\bibitem{BESIII:2009fln}
	M.~Ablikim \textit{et al.} (BESIII Collaboration), \href{https://doi.org/10.1016/j.nima.2009.12.050}{Nucl. Instrum. Meth. A \textbf{614}, 345 (2010)}.

\bibitem{Yu:2016cof}
C.~Yu, \textit{et al.} \href{https://accelconf.web.cern.ch/ipac2016/doi/JACoW-IPAC2016-TUYA01.html}{ Proceedings of IPAC2016, Busan, Korea, 642, 2016}.

\bibitem{Huang:2022wuo}
K.~X.~Huang, Z.~J.~Li, Z.~Qian, J.~Zhu, H.~Y.~Li, Y.~M.~Zhang, S.~S.~Sun and Z.~Y.~You,
		\href{https://doi.org/10.1007/s41365-022-01133-8}{Nucl. Sci. Tech. \textbf{33}, 142 (2022)}.

\bibitem{etof1}
	X.~Li \textit{et al.} \href{https://doi.org/10.1007/s41605-017-0014-2}{Radiat Detect Technol Methods \textbf{1}, 13 (2017)};
	Y.-X. Guo \textit{et al.} \href{https://doi.org/10.1007/s41605-017-0012-4}{Radiat. Detect. Technol. Methods \textbf{1}, 15 (2017)};
	P.~Cao \textit{et al.} \href{https://doi.org/https://doi.org/10.1016/j.nima.2019.163053}{Nucl. Instrum. Meth. A \textbf{953}, 163053 (2020)}.

\bibitem{BESIII:2020eyu}
	M.~Ablikim \textit{et al.} (BESIII Collaboration), \href{https://iopscience.iop.org/article/10.1088/1674-1137/ac1575}{Chin. Phys. C \textbf{45}, 103001 (2021)};\href{https://iopscience.iop.org/article/10.1088/1674-1137/ac84cc}{Chin. Phys. C \textbf{46}, 113003 (2022)}.

\bibitem{luminosities}
  M.~Ablikim \textit{et al.} (BESIII Collaboration), \href{https://iopscience.iop.org/article/10.1088/1674-1137/ac80b4}{Chin. Phys. C \textbf{46}, 113002 (2022)}. 
		The article described the integrated luminosity measurement for data taken at $\sqrt{s}=4.189$, 4.199, 4.209, 4.219, and 4.226 GeV. 
		The integrated luminosity values for the other data samples have been obtained by a similar procedure.

\bibitem{CLEO:2008ojp}
D.~Cronin-Hennessy \textit{et al.} (CLEO Collaboration), 
 \href{https://journals.aps.org/prd/abstract/10.1103/PhysRevD.80.072001}{Phys. Rev. D \textbf{80}, 072001 (2009)}.

\bibitem{GEANT4:2002zbu}
	S.~Agostinelli \textit{et al.} (GEANT4 Collaboration), \href{https://doi.org/10.1016/S0168-9002(03)01368-8}{Nucl. Instrum. Meth. A \textbf{506}, 250 (2003)}.

\bibitem{Jadach:2000ir}
	S.~Jadach, B.~F.~L.~Ward and Z.~Was, \href{https://journals.aps.org/prd/abstract/10.1103/PhysRevD.63.113009}{Phys. Rev. D \textbf{63}, 113009 (2001)}.

\bibitem{Jadach:1999vf}
	S.~Jadach, B.~F.~L.~Ward and Z.~Was, \href{https://doi.org/https://doi.org/10.1016/S0010-4655(00)00048-5}{Comput. Phys. Commun. \textbf{130}, 260 (2000)}.

\bibitem{Lange:2001uf}
	D.~J.~Lange, \href{https://doi.org/10.1016/S0168-9002(01)00089-4}{Nucl. Instrum. Meth. A \textbf{462}, 152 (2001)}.

\bibitem{EVTGEN2}
	R.~G.~Ping, \href{https://doi.org/10.1088/1674-1137/32/8/001}{Chin. Phys. C \textbf{32}, 599 (2008)}.

 \bibitem{Chen:2000tv}
	J.~C.~Chen, G.~S.~Huang, X.~R.~Qi, D.~H.~Zhang and Y.~S.~Zhu, \href{https://journals.aps.org/prd/abstract/10.1103/PhysRevD.62.034003}{Phys. Rev. D \textbf{62}, 034003 (2000)}.

\bibitem{LUNDCHARM2}
	R.~L.~Yang, R.~G.~Ping and H.~Chen, \href{https://iopscience.iop.org/article/10.1088/0256-307X/31/6/061301}{Chin. Phys. Lett. \textbf{31}, 061301 (2014)}.

\bibitem{PHOTOS}
	E.~Richter-Was, \href{https://doi.org/10.1016/0370-2693(93)90062-M}{Phys. Lett. B \textbf{303}, 163 (1993)}.
 
\bibitem{MARK-III:1985hbd}
	R.~M.~Baltrusaitis \textit{et al.} (MARKIII Collaboration), \href{https://journals.aps.org/prl/abstract/10.1103/PhysRevLett.56.2140}{Phys. Rev. Lett. \textbf{56}, 2140 (1986)}.

\bibitem{BESIII:2019jjr}
    M.~Ablikim \textit{et al.} (BESIII Collaboration), \href{https://journals.aps.org/prl/abstract/10.1103/PhysRevLett.123.112001}{Phys. Rev. Lett. \textbf{123}, 112001 (2019)}.


\bibitem{BESIII:2021xox}
	M.~Ablikim \textit{et al.} (BESIII Collaboration), \href{https://link.springer.com/article/10.1007/JHEP06(2021)181}{JHEP \textbf{06}, 181 (2021)}.


\bibitem{Ablikim:2019hff}
	M.~Ablikim \textit{et al.} (BESIII Collaboration), \href{https://iopscience.iop.org/article/10.1088/1674-1137/44/4/040001}{Chin. Phys. C \textbf{44}, 040001 (2020)}.

\end{thebibliography}
\end{document}